\documentclass[prd,amsmath,notitlepage,twocolumn]{revtex4-1}
\usepackage{graphicx}
\usepackage{float}
\usepackage{epstopdf,cancel}
\usepackage{epsf,latexsym,bbm,euscript}
\usepackage{amssymb,amsmath}
\usepackage{mathtools} %used in this file for the command \coloneqq -> :=
\usepackage{times,graphics}
\usepackage{soul,xcolor}
\usepackage{mathtools}

\def\6{{\langle}}
\def\9{{\rangle}}
\newcommand{\defeq}{\vcentcolon=}
\newcommand{\eqdef}{=\vcentcolon}

\newcommand{\be}{\begin{equation}}
\newcommand{\ee}{\end{equation}}
\newcommand{\ba}{\begin{eqnarray}}
\newcommand{\ea}{\end{eqnarray}}

 \newcommand{\mA}{{\mathrm{A}}}

\def\half{{\tfrac{1}{2}}}

\def\pad{{\partial}}

\def\ha{{\hat{a}}}

\def\sg{\textsl{g}}

\def\cO{\mathcal{O}}

\usepackage{url,hyperref}
\hypersetup{colorlinks,linkcolor={blue!55!black},citecolor={red!50!black},urlcolor={blue!45!black}}

\begin{document}

\title{Kerr--Vaidya  black holes}

\author{Pravin Kumar Dahal}
     \affiliation{Department of Physics \& Astronomy,
 Macquarie University, Sydney New South Wales 2109, Australia}

\author{Daniel R. Terno}
\affiliation{Department of Physics \& Astronomy,
 Macquarie University, Sydney New South Wales 2109, Australia}
   \affiliation{Shenzhen Institute for Quantum Science and Engineering,  Department of Physics, Southern University of Science and Technology, Shenzhen 518055, Guandong,  China}

%\email{daniel.terno@mq.edu.au}
\begin{abstract}
Kerr--Vaidya metrics are the simplest nonstationary extensions of the Kerr metric. We explore their properties and compare them with the
near-horizon limits of the spherically symmetric self-consistent solutions (the ingoing Vaidya metric with decreasing mass and the outgoing Vaidya metric with increasing mass) for the evaporating
and accreting physical black holes.
  The Newman--Janis transformation relates the corresponding Vaidya and Kerr-Vaidya metrics. For nonzero angular momentum, the energy-momentum tensor violates the null energy condition (NEC).
However, we show that its structure differs from the standard form of the NEC-violating tensors. The apparent horizon in the outgoing Kerr--Vaidya metric coincides
with that of the Kerr black hole. For the ingoing metric, its location is different. We derive the ordinary differential equation for this surface and locate it numerically. A spherically symmetric accreting black hole
leads to a firewall ---
a divergent energy density, pressure, and flux as perceived by an infalling observer.
We show that this is also true for the outgoing Kerr--Vaidya metric.  \end{abstract}

\maketitle
\section{Introduction}
Black holes are   described both as ``the most perfect macroscopic objects in the universe'' \cite{chandra:b} and as one of the ``most mysterious concepts conceived by the human mind'' \cite{abh}.
Thanks to the successes of the gravitational wave astronomy and direct observations
 of ultracompact objects (UCOs), the old debate about the physical relevance of black hole solutions \cite{israel:86} has been reframed as a question about the nature of UCOs \cite{cp:na17,pheno,bh-map}.

Diversity of opinions about black holes and their significance are matched by absence of a universally accepted definition \cite{curiel}. However, the core idea of a black hole as a spacetime region from which nothing can escape
 is formalized in the notion of a trapped region. Gravity there is so strong that both
ingoing and outgoing future-directed null geodesics originating at a spacelike two-dimensional surface with spherical topology have negative expansion \cite{he:book,fn:book,faraoni:b}.
The apparent horizon is its evolving outer boundary.

A physical black hole   contains such a trapped region \cite{frolov:14}.  To be relevant to distant observers with a finite lifespan it has to be formed in finite time according to their clocks \cite{bmmt:18}.
 Otherwise, black hole solutions can have only approximate or asymptotic meaning. A physical black hole may possess other
 classical features, such as an event horizon and a singularity,
 or be a singularity-free regular black hole. One of the issues at stake   is whether the observed  astrophysical black hole candidates  contain  light-trapping regions, i.e.,
 they are black holes or do not, and thus they are
 horizonless UCOs.

  Quantum effects make the black hole physics particularly interesting \cite{fn:book,faraoni:b,haw:74,bd:82,hv:book,rev-0,rev-1}. On the one hand,  an apparent horizon is accessible to an
observer at infinity  (Bob) only if the classical energy conditions \cite{he:book,exact-b,mmv:17,ks:20}  are violated \cite{he:book}.
The Hawking radiation \cite{fn:book,bd:82,haw:74} has precisely this property. On the other hand, the Hawking radiation precipitates the infamous information loss paradox \cite{rev-0,rev-1}.
One way to resolve the paradox is to have a horizonless UCO or a regular black hole  as the final product of the gravitational collapse \cite{cp:na17,pheno}.
 These objects also require a violation of the energy conditions for their existence.
  Another resolution of the information loss paradox posits that the infalling observer (Alice) does not see a vacuum at the black hole horizon, but instead encounters a large number
   of high-energy modes \cite{rev-0,amps:13}, known as the firewall.

A self-consistent approach \cite{bmmt:18,bmt:18,t:19,t:20} starts with the assumption that physical black holes do form.
 Once the assumption of formation of a singularity-free apparent horizon is translated into mathematical statements, it allows to obtain a number of concrete results.
 In spherical symmetry there are only two possible classes of black hole solutions, and it is possible to identify the amount of violation of the energy conditions that they require. Accreting physical black hole
  solutions %in both classes
lead to divergent energy density, pressure, and flux as experienced by Alice, while the curvature scalars remain finite.

Real astrophysical objects are rotating. Hence %, similarly to other results that were derived from spherically symmetric solutions, 
it is important to verify that
the firewall is not an artifact of the spherical symmetry.
          The Kerr metric is the asymptotic result of the classical collapse \cite{he:book,fn:book,chandra:b}. The simplest models that allow for an axially symmetric variable mass
  distribution are given by the so-called
  Kerr--Vaidya metrics \cite{mt:70,ck:77,st:15}. In Sec.~\ref{sII} we review the relevant properties of the spherically symmetric solutions. In Sec.~\ref{sIII} we discuss their axially symmetric counterparts, 
  focusing on the violation
  of the energy conditions,
   location of the apparent horizons and presence of a firewall. %For two classes of Kerr-Vaidya metrics that may describe a PBH we find the same qualitative features as in the spherically symmetric case.

 \section{Near-horizon regions of spherically symmetric black holes} \label{sII}

Working in the framework of semiclassical gravity \cite{pp:09,bmt-1,hv:book} we use classical notions (horizons, trajectories, etc.),
 and describe dynamics via the Einstein equations $G_{\mu\nu}=8\pi T_{\mu\nu}$, where the Einstein tensor $G_{\mu\nu}$ is
  equated to the expectation value $T_{\mu\nu}=\6\hat{T}_{\mu\nu}\9_\omega$ of the renormalized energy-momentum tensor (EMT).  For simplicity  we consider an asymptotically flat space.
  We do not make any specific assumptions apart from
  (i) the apparent horizon   was formed at some finite time of Bob, (ii)  it is regular, i.e.,  the curvature scalars, such as  $\mathrm{T}\defeq T^\mu_{~\mu}\equiv -\EuScript{R}/8\pi$
   and $\mathfrak{T}\defeq T^{\mu\nu}T_{\mu\nu}\equiv R^{\mu\nu}R_{\mu\nu}/64\pi^2$  are finite at the horizon. (Here $R_{\mu\nu}$ and $\EuScript{R}\defeq R^\mu_{\,\mu}$ are the Ricci tensor and the Ricci scalar, respectively).

      A general spherically symmetric metric   in the  Schwarzschild coordinates  is given by
\be
ds^2=-e^{2h(t,r)}f(t,r)dt^2+f(t,r)^{-1}dr^2+r^2d\Omega, \label{sgenm}
\ee
where $r$ is the areal radius.  The Misner-Sharp mass  \cite{ms,faraoni:b} $M(t,r)$ is invariantly defined  via  $1-2M/r\defeq \pad_\mu r\pad^\mu r$, %\label{defMS}
and   $f(t,r)=1-2M(t,r)/r$. The apparent horizon is located at the Schwarzschild radius $r_\sg$ that is the largest root of
 $f(t,r)=0$ \cite{faraoni:b,aphor}.
%The function $h(t,r)$   may contain   information about potential hairs of the stationary PBHs \cite{fn:book,cch:12}, and plays the role of an integrating factor in the coordinate transformations.

Only two near-horizon forms of the EMT and the metric  are consistent with the above two assumptions. Here we consider the generic form that agrees with the \textit{ab initio}
  calculations of the EMT on the background of the Schwarzschild solution \cite{leviori:16}.  In this case
the leading terms in the expansion of the  metric functions in terms of $x\defeq r-r_\sg(t)$ are
          \begin{align}
&2M(t,r)= r_\sg-w\sqrt{x}+\cO(x),\\
 &h(t,r)=-\frac{1}{2}\ln{\frac{x}{\xi}}+\cO\big(\sqrt{x}\big),  \label{k0met}
\end{align}
where the function  $\xi(t)$ is   determined by the  choice of the time variable (and requires for its determination knowledge of the full solution of the Einstein equations), and
$w^2\defeq 16 \pi \Upsilon^2 r_\sg^3$  characterizes the leading behavior of the EMT \cite{bmmt:18}.

In particular,
in the  orthonormal basis the $(\hat t\hat r)$  block of the EMT near the apparent horizon is given by
    \be
T_{\hat a \hat b}=-\frac{\Upsilon^2}{f} \!\!\begin{pmatrix}
1& \pm 1 \\
\pm 1 & 1 \end{pmatrix}.
\ee
 The upper (lower) signs of $T_{\hat t \hat r}$ correspond to evaporation (growth) of the physical black hole. Consistency of the Einstein equations  results in the relation
 \be
 r'_\sg/\sqrt{\xi}=\pm4\sqrt{\pi}\,\Upsilon\sqrt{ r_\sg}= \pm w/r_\sg.      \label{lumin}
 \ee
     The null energy condition (NEC)   requires  $T_{\mu\nu} l^\mu l^\nu\geqslant0$ for all null vectors $l^\mu$ \cite{he:book,mmv:17,ks:20}. It is violated
     by radial vectors $l^{\hat a}=(1, \mp 1,0,0)$ for the evaporating and the accreting solutions, respectively \cite{bmmt:18} .

Null coordinates allow to represent the near-horizon geometry in a simpler form.     The advanced null   coordinate $v$,
 \be
dt=e^{-h}(e^{h_+}dv- f^{-1}dr), \label{intf}
\ee
is useful in the case   $r'_\sg<0$. A general spherically symmetric metric in $(v,r)$ coordinates is   given by
\be
  ds^2=-e^{2h_+}\left(1-\frac{C_+}{r}\right)dv^2+2e^{h_+}dvdr +r^2d\Omega. \label{lfv}
  \ee
 Using the Einstein equations and the relationships between components of the EMT in  two coordinates systems \cite{t:20} one can show that
\begin{align}
&C_+(v,r)=r_+(v)+w_1(v) x +\ldots, \\
&h_+(v,r)=\chi_1(v) x +\ldots,          \label{Vaidyav}
\end{align}
where   $r_+(v)$ is the radial coordinate of the apparent horizon, $C_+(v,r_+)\equiv r_+$, $x\eqdef r-r_+(v)$.%, and the functions $w_2$ and $\chi_2$ are related to the higher-order terms in the EMT.
As a result, at the apparent horizon both the metric  corresponds to the Vaidya geometry with $C'_+(v)=2M'(v)<0$.

If  $r'_\sg>0$  it is useful to switch to the retarded null coordinate $u$.
The near-horizon geometry is then described by  the Vaidya metric with $C_-'(u)>0$.

A static observer finds that the energy density $\varrho=T_{\mu\nu}u^\mu u^\nu=-T^t_{~t}$, the pressure $p=T_{\mu\nu}n^\mu n^\nu=T^r_{~r}$, and the
flux $\phi\defeq  T_{\mu\nu}u^{\mu}n^\nu$  (where $u^\mu$ is the four-velocity and $n^\mu$ is the outward-pointing radial spacelike vector),
 diverge at the apparent horizon.   A radially infalling Alice moves on a trajectory   $x^\mu_\mathrm{A}(\tau)=(T(\tau), R(\tau),0,0)$. Horizon crossing
happens not only at some finite proper time $\tau_0$,  $r_\sg\big(T(\tau_0)\big)=R(\tau_0)$, but thanks to the form of the metric also at a finite time $T(\tau_0)$ of Bob.

 However, experiences of Alice are  different at the apparent horizon of an evaporating and accreting physical black holes. For an evaporating black hole, $r_\sg'<0$, energy density, pressure and flux are finite.
  For example, if we approximate the near-horizon geometry by a pure outgoing Vaidya metric with $M'(v)<0$, Alice's energy density at the horizon crossing is
 \be
 \varrho_{\mathrm{A}}^<=p_{\mathrm{A}}^<=\phi_{\mathrm{A}}^<=-\frac{\Upsilon^2}{4\dot R^2},            \label{comov}
 \ee
at $r_\sg=R$.
For an accreting black hole, $r_\sg'>0$, Alice experiences
  the divergent values of energy density, pressure and flux,
  \be
\varrho_{\mathrm{A}}^>=p_{\mathrm{A}}^>=-\phi_{\mathrm{A}}^>=-\frac{2\dot R^2\Upsilon^2}{F^2 }+\cO (F^{-1}),     \label{pdiverp}
\ee
in the vicinity of the apparent horizon, as $F\defeq f(T,R)\to 0$.

Thus an expanding trapped region is accompanied by a  firewall--- a region of unbounded energy density, pressure, and flux---that is perceived by an infalling observer. Unlike the
  firewall from the eponymous paradox, it appears as a consequence of regularity of the expanding apparent horizon and its finite formation time.
                                     The divergent energy density leads to a violation \cite{t:19,t:20} of the inequality that bounds the amount of negative energy along a timelike trajectory in a moderately curved
                                      spacetime \cite{ko:15}.   As a result a physical black hole, once formed, can only evaporate. Another possibility is that
the semiclassical physics breaks down at the horizon scale.

 \section{Kerr--Vaidya metric}    \label{sIII}
      A general time-dependent axisymmetric metric contains seven  functions of three variables (say, $t$, $r$ and $\theta$) that enter the Einstein equations via six independent combinatons \cite{chandra:b}. However, to verify
that certain predictions of the self-consistent approach (such as a firewall at an expanding apparent horizon)   are not  an artefact of the spherical symmetry, it is enough to consider a simpler geometry.

 The Kerr metric can be represented      using either the ingoing   \cite{poisson:b}
 \begin{align}
    & ds^2=-\bigg(1-\frac{2 M r}{\rho^2}\bigg)dv^2+2 dv dr-
    \frac{4 a M r \sin^2\theta}{\rho^2} dvd\psi    \nonumber \\
   & -2 a \sin^2\theta dr d\psi  \!  +\rho^2 d\theta^2   +
    \frac{(r^2+a^2)^2\!-\!a^2 \Delta \sin^2\theta}{\rho^2}\sin^2\theta d\psi^2,
    \label{kv}
 \end{align}
 or the outgoing  null congruences \cite{chandra:b},
  \begin{align}
    & ds^2=-\bigg(1-\frac{2 M r}{\rho^2}\bigg)du^2-2 du dr -      \frac{4 a M r \sin^2\theta}{\rho^2}du d\psi   \nonumber \\
   & +2 a \sin^2\theta d\psi dr \!   +\rho^2 d\theta^2   +
    \frac{(r^2+a^2)^2\!-\!a^2 \Delta \sin^2\theta}{\rho^2}\sin^2\theta d\psi^2,
    \label{ku}
 \end{align}
  where $\rho^2\defeq r^2+a^2 \cos^2\theta$, $\Delta\defeq r^2-2 M r+a^2$, and $a=J/M$ is the angular momentum per unit mass.

The easiest way to obtain this result is to follow the complex-valued  Newman--Janis  transformation \cite{nj:65} starting with the Schwartzschild metric written in the ingoing or the outgoing Eddington--Finkelstein coordinates.
The simplest nonstationary generalizations  of the Kerr metric  are obtained by introducing evolving  {masses} $M(v)$ and $M(u)$.
The metric of Eq.~\eqref{ku} with a variable $M(u)$ is obtained from  the retarded Vaidya metric \cite{heje:79,gm:15}.   By using
   the advanced Vaidya metric of Eq.~\eqref{lfv} as the seed metric, the metric  of Eq.~\eqref{kv} can be obtained following the same procedure (Appendix~\ref{aA}).
   
  \subsection{Energy conditions}

 %It is important to verify that the exotic matter of Eq.~\eqref{comov}  and the   firewall of Eq.~\eqref{pdiverp} are not artefacts of the spherical symmetry.

   % As with the Vaidya metric, four types of Kerr-Vaidya metric are possible, but only two are compatible with the finite formation time according to Bob.
A schematic form of the EMT in both cases is
\be
T_{\mu\nu}=\begin{pmatrix}
T_{oo}& 0 & T_{o\theta} &T_{o\psi} \\
0 &0 &0 & 0 \\
T_{o\theta} & 0 &0 & T_{\theta\psi} \\
T_{o\psi} & 0 &T_{\theta\psi} & T_{\psi\psi} \end{pmatrix},\label{emt}
\ee
where $o=u,v$. Using the null vector $k^\mu=(0,1,0,0)$ \cite{mt:70} the EMT can be represented as
\be
T_{\mu\nu}=T_{oo}k_\mu k_\nu+q_\mu k_\nu+q_\nu k_\mu, \label{MT-emt}
\ee
where the components of $T_{\mu\nu}$ and of the auxiliary vector $q_\mu$, $q_\mu k^\mu=0$, for both cases are given in Appendix~\ref{aB}. The EMT [for the metric Eq.~\eqref{ku}] was identified in Ref.~\cite{cgk:90} as
 belonging to the type $[(1,3)]$ in the Segre classification \cite{exact-b}, i.e., to the type III of the Hawking--Ellis classification \cite{he:book,mmv:17}, indicating that the NEC is violated for any $a\neq 0$.

A detailed investigation reveals some interesting properties of this EMT. We use a tetrad in which the null eigenvector $k^\mu=k^{\ha}e^\mu_\ha$ has the components $k^\ha=(1,1,0,0)$,
the third vector $e_{\hat 2}\propto \pad_\theta$ and the remaining vector $e_{\hat{3}}$ is found by completing the basis, the EMT takes the form
 \be
T^{\ha\hat{b}}=\left(\begin{tabular}{cc|cc}
 $\nu$ & $\nu$  & $q^{\hat 2}$& $q^{\hat 3}$\\
 $\nu$ & $\nu$  & $q^{\hat 2}$& $q^{\hat 3}$\\ \hline \label{IVP}
 $q^{\hat 2}$& $q^{\hat 2}$ & 0 & 0 \\
$q^{\hat 3}$ & $q^{\hat 3}$& 0 & 0
 \end{tabular}\right).
\ee
Explicit expressions for  the tetrad vectors and the matrix elements are given in Appendix \ref{aB}.

 For an arbitrary null vector   $l_\ha=(-1,n_\ha)$,
$n_\ha=(\cos\alpha,\sin\alpha\cos\beta,\sin\alpha\sin\beta)$ the NEC  becomes
\be
\nu(1-\cos\alpha)+2\sin\alpha({q^{\hat 2}} \cos\beta+q^{\hat 3}\sin\beta)\geqslant 0.
\ee 
This inequality is satisfied if and only if $\nu\geqslant 0$ and $q^{\hat 2}=q^{\hat 3}=0$. 
The condition $q^{\hat 2}=q^{\hat 3}=0$ holds only when $a=0$, so the metric reduces to its Vaidya
 counterpart and the EMT becomes a type II tensor. Only in this case  the
NEC may be satisfied.

Each type of the EMT is characterized by its Lorentz-invariant eignevlaues   \cite{exact-b,mmv:17}. These are the eigenvalues of the matrix $T^{\ha}_{~~\hat{b}}$, i.e., 
 the roots of the equation
\be
 \det (T^{\ha\hat{b}}-\lambda \eta^{\ha\hat{b}})=0, \qquad \eta^{\ha\hat{b}}=\mathrm{diag}(-1,1,1,1).
\ee
The EMT of Eq.~\eqref{IVP} has a single quadruple-degenerated Lorentz-invariant eigenvalue $\lambda=0$.    On the other hand, two of the eigenvalues of the matrix $\tilde\lambda$ of $T^{\ha\hat{b}}$ are nonzero,
\be
\tilde\lambda_{1,2}=\nu\pm\sqrt{2( q_{\hat 2}^2+q_{\hat 3}^2)+\nu^2}. 
\ee
As a result the EMT tensor \eqref{IVP} cannot be brought to a generic type III
   form by an arbitrary similarity transformations  unless $\tilde\lambda_1=-\tilde\lambda_2$, which is impossible
for $M'\neq 0$ (see Appendix~\ref{aB} for the details).

\subsection{Apparent horizon}

The apparent horizon of the Kerr black hole coincides with its event horizon. It is located at the largest
  root of $\Delta=0$,
  \be
  r_0\defeq M+\sqrt{M^2-a^2}.
  \ee
             For  both the ingoing and the outgoing  Vaydia metrics the apparent horizon is located at $r_\sg=r_0=2M$. For the  metric \eqref{ku} the relation $r_\sg=r_0$ also holds \cite{x:99},
              but it fails for the metric \eqref{kv} \cite{st:15}.
             In this case the
 difference $r_\sg(v,\theta)-r_0(v)$ is of the order $|M_v|$.

 We now identity its location  in the foliations with the hypersurfaces $v=\mathrm{const}$.
The standard approach \cite{nr,tho:07} for constructing the ordinary differential equation for the apparent horizon is based on exploiting properties of a spacelike foliation.  It cannot be used in this case as the
foliating hypersurfaces are timelike. However, since the approximate location of the apparent horizon is known,
we obtain the leading correction in $M_v$ by using the  methods of analysis of null congruences and hypersurfaces \cite{poisson:b}.
%  and adapting the approach used for the numerical search of axially symmetric apparent horizons.

   \begin{figure}[htbp]
\includegraphics[width=0.45\textwidth]{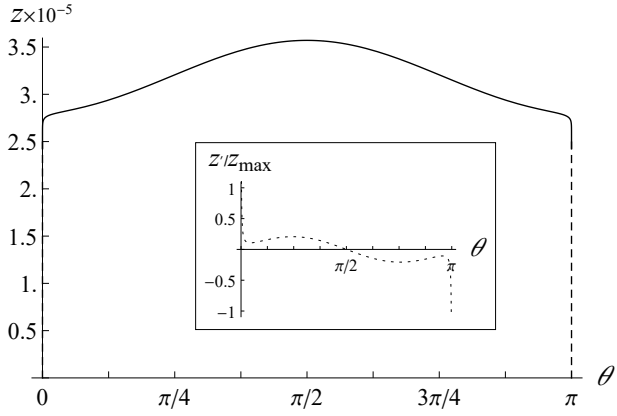}
\caption{{Location of the apparent horizon relative to $r_0$ for $M=1$, $a=0.1$, $M_v=-\kappa/M^2=0.01$}. The equation was first solved as a boundary value problem $z(0)=0$, $z'(\pi/2)=0$,
resulting in $z_\mathrm{m}\defeq z(\pi/2)\approx3.57\times 10^{-5}$. Solution of the initial value problem $z(\pi/2)=z_\mathrm{m}$, $z'(\pi/2)=0$ coincides with the previous one
within the relative precision of $2\times 10^{-15}$ outside $\delta=10^{-6}$ interval from the poles.  }
\label{schema1}
\end{figure}
        
Assume that at some advanced time $v$ the apparent horizon $S_0(v)$ is located  at $r_\sg=r_0(v)+z(\theta)$, where the function $z(\theta)$ is to be determined.
Once the future-directed outgoing null geodesic congruence orthogonal the surface is
identified, calculating the expansion $\vartheta$ and equating it to zero results in the differential equation for $z(\theta)$.  There are at least two equivalent ways to obtain this equation.

The outward and inward-pointing null vectors  $l^\mu$ and $N^\mu$, respectively, are defined on $S_0$. They are   orthogonal to its tangents and can be normalized by  the condition $N_\mu l^\mu=-1$.
These vectors can be extended to the fields of tangent vectors to the families of affinely parametrized null geodesics in the bulk.

One approach to calculation of the expansion uses its geometric meaning as a relative rate of change of the two-dimensional cross section area   \cite{he:book,poisson:b}. 
 Consider an infinitesimal geodesic triangle $(x_\mathrm{in} x_\theta x_\psi)$  
 on the surface $S_0$. It is defined by three vertices
   \be
x_\mathrm{in}^\mu, \qquad x_\psi^\mu=x^\mu_\mathrm{in}+b^\mu_\psi \delta\psi,     \qquad x_\theta^\mu=x^\mu_\mathrm{in}+b^\mu_\theta\delta\theta.       \label {2point}
  \ee
    The two tangent vectors
    \be
 \left.   b^\mu_\psi\defeq \frac{\pad x^\mu}{\pad\psi}\right|_{S_0}, \qquad \left. b^\mu_\theta\defeq \frac{\pad x^\mu}{\pad \theta}\right|_{S_0},
    \ee
     introduce a metric two-tensor
  \be
  \sigma_{AB}\defeq \sg_{\mu\nu}b^\mu_A b^\nu_B, \qquad A,B=\theta, \psi.     \label{met2d}
  \ee
The area of the  triangle $(x_\mathrm{in} x_\theta x_\psi)$ is $\delta A=\half \sqrt\sigma\delta\theta\delta\psi$, where   $\sigma =\sigma_{11}\sigma_{22}-\sigma_{12}^2$ is the determinant of the two-dimensional metric.
Under the geodesic flow $x^\mu\to x^\mu+l^\mu(x)d\lambda$ the coordinates  $\theta$ and $\psi$  are comoving and thus constant for the three vertices, but both the metric $\sg$ and the vectors $b_A$ evolve with $\lambda$.

Calculating the ratio of the first-order area change $\delta A(\lambda+d\lambda)-\delta A(\lambda)$  to the initial area of the triangle allows to obtain
the expansion by using the relation 
\be
 \vartheta=\frac{1}{\sqrt{\sigma}}\frac{d\sqrt{\sigma}}{d\lambda},
 \ee
  where $\lambda$ is the affine parameter. Solution of the second-order differential equation $\vartheta(z)=0$ gives the desired function $z(\theta)$.
  An alternative derivation is based on the direct evaluation of  $\vartheta=l^\mu_{;\mu}$ on $S_0$
 (see Appendix \ref{aC} for the details).

 If both $z(\theta)$ and its derivatives are much smaller than $r_0$, we obtain a linear ordinary differential equation  for $z$, \begin{widetext}
 \be
 8 r_0^2(a^2+r_0^2)^2(  z''+\cot \theta z') -(r_0^2-a^2)\big (a^4+7 a^2 r_0^2 +8 r_0^4+a^2(r_0^2- a^2)  \cos 2 \theta \big)z - 8 r_0^3a^2(a^2+r_0^2) \sin ^2 \theta M_v=0,   \label{odeZ}
\ee
                      \end{widetext}
 where the regular singular term     $(\cot\theta z'+z'')$ is a standard feature of the apparent  horizon     equations \cite{nr,tho:07}. Symmetry considerations lead to the condition $z(0)=z(\pi)=0$.  
  The second initial condition  $z'(\pi)=0$ ensures that the surface of the apparent horizon is smooth \cite{nr}. 

 A typical result of the numerical solution is depicted in Fig.~\ref{schema1}. It was obtained by imposing the boundary conditions $z(0)=0$ and
 $z'(\pi/2)=0$, that enforces the equatorial symmetry.
 The assumption of $|z'|\ll r_0$ fails near the poles, where $z=0$. This is not an artefact of the approximation.
 Using the series solutions of Eq.~\eqref{odeZ}  with the conventional initial conditions
  $z(0)=0$, $z'(0)$=0  \cite{nr,tho:07}, i.e., in the regime where the assumption $|z'|\ll r_0$ is clearly valid,  leads to
  \be
  z_\mathrm{ser}= \frac{a^2 M_v r_0}{16(a^2+r_0^2)}\theta^4+\cO(\theta^5).
  \ee
  For $M_v>0$ it implies that at least near the poles $r_\sg>r_0$, i.e., at $r=r_0$ the expansion is still negative.
   However, this is impossible: at the poles the null congruence that is orthogonal to the two-dimensional surface $r=r_0$ \cite{st:15} has   $\vartheta>0$.
     Moreover, using this solution to provide the initial values $z(\theta)$, $z'(\theta)$ at some
    $\theta=\epsilon\ll 1$  leads to inconsistencies.

We investigated stability of this result in numerical experiments. For a fixed $M_v=-\kappa/M^2$ the initial value problem $z(\pi/2)=z_0$,  $z'(\pi/2)=0$, where $z_0$ is some number,
leads to a well-behaved numerical solution.
However, the conditions $z(0)=z(\pi)=0$    are satisfied within a prescribed tolerance only for a very narrow range of the values $z_0$ around $z_\mathrm{m}=z(\pi/2)$ of the numerical solution of the above
 boundary value problem.
We will provide a full analysis of the apparent horizon in  a future work.

 \subsection{Firewall}

All components of the EMT \eqref{emt} are finite at the apparent horizon. Divergences of the comoving parameters can appear only as a result of divergences in the components of the four-velocity of Alice.
We now show that similarly to the spherically symmetric geometries density and pressure in Alice's frame are finite if Alice crosses the apparent horizon in the metric of Eq~\eqref{kv},
 but diverge for the metric of Eq.~\eqref{ku}.

  In the spherically symmetric case  Alice was a  zero angular momentum observer (ZAMO) \cite{fn:book, poisson:b}. In axially symmetric spacetimes
the ZAMO condition results in a nontrivial angular velocity $\Psi_Z$.  We begin with the retarded Kerr--Vaidya metric,   where the apparent horizon is located at $r_\sg=r_0$. Alice's four-velocity is
   \be
 u^\mu_\mathrm{A}=\big(\dot{U},  \dot{R},  \dot{\Theta},\dot \Psi_Z\big),
 \ee
where the ZAMO condition $\xi_\psi \cdot u_\mathrm{A}=0$, where  the Killing vector $\xi_\psi=\pad_\psi$, implies $\dot\Psi_Z=-(\sg_{u\psi}\dot U+\sg_{r\psi}\dot R)/\sg_{\psi\psi}$.
During the fall $\dot R<0$. The velocity component $\dot U>0$
is obtained from the normalization condition $u_\mathrm{A}^2=-1$,
 \be \dot U=  -\frac{\dot R^2}{\Delta}(r^2+a^2) +  \frac{1}{\Delta\rho}
\sqrt{(\Delta(1+\rho^2\dot \Theta^2)+\rho^2 \dot R^2)\Sigma}\,, \ee where $\Sigma=(a^2+r^2)\rho^2+2 a^2 r M \sin^2\theta$.
 As $X\defeq R(\tau)-r_0\big(U(\tau)\big)\to 0$
 the derivative $\dot U$  diverges as $\Delta^{-1}$,
 \be
 \dot U= -\frac{2 r_0 \dot R M}{X(r_0-M)}+\cO(X).
 \ee

The energy density in Alice's frame is given then by
\be
\rho_\mathrm{A}=\left( T_{uu}+T_{\psi\psi}\left(\frac{g_{u\psi}}{g_{\psi\psi}}\right)^2-2T_{u\psi}\frac{g_{u\psi}}{g_{\psi\psi}}\right) \dot U^2+\cO\big(\Delta^{-1}\big),
\ee
resulting in
\begin{align} \rho_\mathrm{A}&\approx \frac{(-2M'-(2M-r_0)\sin^2\theta M'')r_0^2\dot R^2}{8\pi X^2(r_0-M)^2}\nonumber \\
 &= \frac{(-2 r_0 M'-a^2 \sin^2\theta M'')r_0 \dot R^2}{8\pi X^2(r_0-M)^2}.
\end{align}
We choose the spacelike direction analogously to the spherically symmetric case,
\be
n^\mathrm{A}_\mu=(-\dot R, \dot U,0,0).
\ee
Then (after setting $\dot \Theta=0$),
\be
p_\mathrm{A}=T_{\mu\nu}n_\mathrm{A}^\mu n_\mathrm{A}^\nu   \approx \frac{(-2r_0M'{-}a^2\sin^2\theta M'')r_0\dot R^2}{8\pi X^2(r_0-M)^2}. \ee
It is easy to see that for $a=0$ we recover the firewall of the outgoing Vaidya metric.

Violations of the NEC are bounded by quantum energy inequalities (QEIs) \cite{ks:20,few:17}. For spacetimes of small curvature explicit expressions that bound time-averaged energy density for a geodesic observer
  were derived in Ref.~\cite{ko:15}.
   For any Hadamard state $\omega$ and a sampling
function $\mathfrak{f}(\tau)$ of compact support,
 negativity of the expectation  value of the energy density  $\rho=\6 \hat{T}_{\mu\nu}\9_\omega u^\mu u^\nu$ as seen by a geodesic observer that moves on a trajectory $\gamma(\tau)$ is bounded by
\be
 \int_\gamma\! \mathfrak{f}^2(\tau)\rho d\tau \geqslant - B(R,\mathfrak{f},\gamma), \label{qei}
\ee
where $B>0$ is a bounded function that depends on the trajectory, the Ricci scalar and the sampling function \cite{ko:15}.

Consider a growing apparent horizon, $r_0'(u)>0$. For simplicity we consider a polar trajectory $\theta=0$.
 For a macroscopic black hole the curvature at the apparent
 horizon is low and thus Eq.~\eqref{qei} is applicable.  Horizon radius (and mass, as in this model $a=\mathrm{const}$), do not appreciably change while
  Alice moves in its vicinity.  Hence $dM/d\tau=M'(U)\dot U\approx \mathrm{const}$ and $\dot X \approx \dot R$. Given Alice's
 trajectory we can choose  $\mathfrak{f}\approx 1$ at the horizon crossing and $\mathfrak{f}\to 0$  within the NEC-violating domain (as Eq.~\eqref{kv} can be valid only in the vicinity of the horizon). As the trajectory
 passes through  $X_0+r_\sg\to r_\sg$ the lhs of Eq.~\eqref{qei} behaves as
         \begin{align}
         & \int_\gamma\! \mathfrak{f}^2\rho_\mA d\tau\approx -\int_\gamma\frac{M'r_0^2\dot R^2d\tau}{4\pi   X^2(r_0-M)^2} \nonumber \\
& \approx {\int_\gamma\frac{M_\tau r_0\, dX}{8\pi  M (r_0-M)X}}\propto \log X_0\to-\infty,
 \end{align}
  where $M_\tau=M'\dot U$ and we used $\dot R\sim\mathrm{const}$. The rhs  of Eq.~\eqref{qei} remains finite, and thus the QEI is violated.  This violation indicates {that} the apparent horizon cannot expand, similarly to
  the spherically symmetric case.

  On the other hand, nothing dramatic happens to the comoving density and pressure in the ingoing Kerr--Vaidya metric. Following the same steps we find that, e.g., the comoving energy density  for the motion in the
equatorial plane ($\Theta=\pi/2$, $\dot \Theta=0$), is
\be
\rho_\mathrm{A}= \frac{T_{vv}}{4\dot R^2}+\cO(a^2). \ee
This quantity is finite and for $a=0$   reduces to Eq.~\eqref{comov}.

  \section{Discussion}
 Extending the self-consistent approach of horizon analysis to the axially symmetric spacetimes is difficult.%, as the most general axially symmetric metric in four spacetime dimensions contains seven
%  functions that are subject only to one constraint \cite{chandra:b}.   
 Kerr--Vaidya metrics are the simplest nonstationary extension of the Kerr
solution. All Kerr--Vaidya metrics violate classical energy conditions. While it could have been previously considered as a drawback, this violation is a
 necessary condition to describe an object with a trapped region that is accessible, even   in principle,  to a distant observer.
   Moreover,  Kerr--Vaidya metrics are related by the Newman--Janis transformation to the pure Vaidya metrics that describe the geometry of physical black holes near their apparent horizons.

 These simple geometries have several remarkable properties. The EMT of the Kerr-Vaidya metric, while violating the NEC for all $a\neq 0$ is a special case of type III form of the EMT in the
 Segre--Hawking--Ellis classification.
  An expanding spherically symmetric apparent horizon leads to a firewall and violates the quantum energy inequality that bounds the amount
of negative energy in spacetimes of low curvature.  The outgoing Kerr-Vaidya metric has the same property, showing that the firewall is not an artifact of spherical symmetry.

The apparent horizon of the
 outgoing Kerr-Vaidya metric coincides with the event horizon $r_0=M+\sqrt{M^2-a^2}$ of the Kerr metric, $M(u)=\mathrm{const}$. For the  ingoing the two surfaces are different. However, the  difference
 $z(\theta)=r_\sg-r_0$ is small if $|M'(v)|\ll 1$, as in this case  $z\propto M'$.  However, while at the poles $z(0)=z(\pi)=0$,  a commonly used assumption  $z'(0)=0$ does not hold. As a result, the apparent horizon is not
 a smooth surface.

 The assumption $a=\mathrm{const}$
is incompatible with the continuous eventual evaporation of a physical black hole, as for $M <a$ the equation $\Delta=0$ has no real roots and the Hawking temperature
\be
T=\frac{1}{2\pi}\left(\frac{r_0-M}{r_0^2+a^2}\right),
\ee
that is proportional to the surface gravity, goes to zero as $M\to a$. Moreover, the semiclassical analysis \cite{fn:book}  shows that during evaporation $a/M$ decreases faster than $M$  \cite{dks:16,aas:20}.

The variability of $a=J/M$ ratio
should not affect existence of the firewall for accreting PBHs, as it is exhibited as a result $\Delta\to 0$ effect in $(vr)$ coordinates and holds for $a=0$. We will drop the assumption $a=\mathrm{const}$
in the future work, and will
 to use the-self consistent approach to match the semiclassical results \cite{dks:16,aas:20,leom:17}, as it was done in
the spherically symmetric case. \acknowledgments

The work of PKD is supported by IMQRES. Useful discussions with Pisin Chen, Eleni Kontou and Sebastian Murk, and helpful comments of Luis Herrera, Joey Medved  and AJ Terno are gratefully acknowledged.

%%%%%%%%%%%%%%%%%%%%%%%%%%%%%%%%%%%%%%%%%%%%%%%%%%%%%%%%%%%%%%%%%%%%%%%
%%######################################################################
\appendix

\section{The Newman--Janis\\
transformation of the advanced Vaidya metric} \label{aA}

The procedure follows the Newman-Janis prescription \cite{nj:65,gm:15} that is applied to the Vaidya metric in advanced coordinates as the seed metric. We use the null tetrad
that consists of a pair of real \cite{chandra:b} 
\begin{align}
& l^\mu=\delta^\mu_v+ \frac{1}{2}f(v,r) \delta^\mu_r, \\
& n^\mu=-\delta^\mu_r 
\end{align}
and a pair of complex-conjugate vectors
\be
  m^\mu=\frac{1}{\sqrt{2}r}\left(\delta^\mu_\theta+\frac{i}{\sin\theta}\delta^\mu_\psi \right), \qquad {\bar m}^\mu=(m^\mu)^*,
\ee that satisfy the standard completeness and orthogonality relations,
\begin{equation}
    \begin{split}
     &   l^\mu l_\mu =l^\mu m_\mu =l^\mu {\bar m}_\mu=0,\\
     &   n^\mu n_\mu=n^\mu m_\mu =n^\mu {\bar m}_\mu=  m^\mu m_\mu=0,\\
     &   l^\mu n_\mu=  -m^\mu {\bar m}_\mu=-1.
    \end{split}
\end{equation}
The metric
\begin{equation}
    ds^2=-f(v,r) dv^2+2 dv dr+r^2 d\theta^2 +r^2 \sin^2\theta d\psi^2,
\end{equation}
where $f(v,r)=1-{2 M(v)}/{r}$, is rewritten as \begin{equation}
    \sg^{\mu\nu}=- l^\mu n^\nu- l^\nu n^\mu+ m^\mu {\bar m}^\nu+ m^\nu {\bar m}^\mu.
\end{equation}
We treat $r$ and $v$ is  complex-valued coordinates  and introduce a real-valued   function
  \be
  f=1- M\big(\half(v+ v^*)\big) \left(\frac{1}{r}+\frac{1}{  r^*}\right),
  \ee
that coincides with $f(v,r)$ for real values of the coordinates, $v=v^*$, $r=r^*$. The   complex coordinate transformation
\begin{equation}
    x'^\mu=x^\mu- i a(\delta^\mu_r+\delta^\mu_v )\cos\theta,
\end{equation}
i. e.,
\begin{align}
 & v'=v-i a \cos\theta, \qquad  \theta'=\theta, \\ & r'=r-i a \cos\theta,  \qquad \psi'=\psi,
\end{align}
 leaves $M$ invariant and transforms the tetrad as
     \begin{align}
& l'^\mu=\delta^\mu_v+ \frac{1}{2}\EuScript{F} (v,r, \theta) \delta^\mu_r, \qquad n'^\mu=-\delta^\mu_r, \\
&m'^\mu=\frac{1}{\sqrt{2}(r-i a \cos\theta)}\left(i a \left(\delta^\mu_v+ \delta^\mu_r \right)\sin\theta +\delta^\mu_\theta+ \frac{i}{\sin\theta} \delta^\mu_\psi \right),
     \end{align}
     where after restricting to the real-valued coordinates
\be
\EuScript{F}=1-2M(v)r/\rho^2.
\ee
Substituting these explicit expressions into the transformed metric \begin{equation}
    \sg'^{\mu\nu}=- l'^\mu n'^\nu- l'^\nu n'^\mu+ m'^\mu {\bar m}'^\nu+ m'^\nu {\bar m}'^\mu,
\end{equation}
produces the Kerr-Vaidya metric in the advanced coordinates that is given in Eq.~\eqref{kv}.

\section{Energy-momentum tensor and the NEC\\ violation for Kerr--Vaidya metric}    \label{aB}

The nonzero components of the energy-momentum tensor for the Kerr--Vaidya metric in advanced coordinates are
                      \begin{align}
               &  T_{v v}=\frac{ r^2(a^2+ r^2)- a^4 \cos^2\theta\sin^2\theta}{4\pi\rho^6}M_v
        -\frac{a^2 r \sin^2\theta}{8\pi\rho^4}M_{vv},\\
                                                                 & T_{v \theta}=-\frac{ a^2 r \sin\theta \cos\theta}{4\pi \rho^4}M_v,\\
                         & T_{v \psi} = - a \sin^2\theta T_{v v}-a \sin^2\theta\frac{r^2-a^2\cos^2\theta}{8\pi\rho^4}M_v, \\
        & T_{\theta \psi}=\frac{ a^3 r \sin^3\theta \cos\theta}{4\pi \rho^4}M_v,\\
         & T_{\psi \psi}= a^2 \sin^4\theta T_{v v}+ a^2 \sin^4\theta\frac{r^2-a^2\cos^2\theta}{4\pi\rho^4}M_v. \end{align}
The nonzero components of the energy-momentum tensor for the Kerr--Vaidya metric in retarded coordinates are
\begin{align}
            & T_{u u}=-\frac{r^2( a^2 +r^2)- a^4 \cos^2\sin^2\theta}{4\pi\rho^6}M_u
        -\frac{a^2 r \sin^2\theta}{8\pi\rho^4}M_{u u}, \label{c4}\\
        & T_{u \theta}=-\frac{2 a^2 r \sin\theta \cos\theta}{8\pi \rho^4}M_u,\\
        &  T_{u \psi}=-  a\sin^2\theta T_{u u}+a\sin^2\theta \frac{r^2-a^2\cos^2\theta}{8 \pi\rho^4}M_u,\\
        & T_{\theta \psi}=\frac{2 a^3 r \sin^3\theta \cos\theta}{8\pi \rho^4}M_u,\\
        &  T_{\psi \psi}= a^2 \sin^4\theta T_{u u}-a^2 \sin^4\theta\frac{(r^2-a^2\cos^2\theta)}{4\pi\rho^4}M_u. \end{align}

In the advanced coordinate the decomposition \eqref{MT-emt} of the EMT is obtained with the vectors
\be
k_\mu=(1,\; 0,\; 0,\; -a\sin^2\theta), \
\ee
and
\be
q_\mu=\bigg(0,\; 0,\; T_{v \theta},\; -a\sin^2\theta\frac{r^2-a^2\cos^2\theta}{8\pi\rho^4}M_v\bigg).
\ee The orthonormal tetrad with where $k^\mu=e_{\hat 1}^\mu+e_{\hat 0}^\mu$ is given by
\begin{align}
&e_{\hat 0}^\mu= (-1,rM/\rho^2,0,0),    \\
&e_{\hat 1}^\mu= (1,1-rM/\rho^2,0,0)  \\
&e_{\hat 2}^\mu=(0,0,1/\rho,0) \\
&e_{\hat 3}^\mu=\frac{1}{\rho}\big(a\sin\theta,a\sin\theta,0,\csc\theta\big).
\end{align}

Hence the EMT is given by Eq.~\eqref{IVP} with     $\nu=T_{vv}$ and $q^\mu=q^{\hat 2} e_{\hat 2}^\mu+q^{\hat 3} e_{\hat 3}^\mu$ with
\begin{align} q^{\hat 2}=& -\frac{a^2 r M_v}{8\pi \rho^5} \sin2\theta, \\
q^{\hat 3}=&{ -\frac{a(r^2 - a^2\cos^2\theta) M_v}{8\pi \rho^5}\sin\theta.}
\end{align}

A generic form \cite{mmv:17} of a type III EMT is
 \be
T^{\ha\hat{b}}=\left(\begin{tabular}{ccc|c}
 $\varrho$ & $0$  & $\varphi$& $0$\\
 $0$ & $-\varrho$  & $\varphi$& $0$\\
 $\varphi$& $\varphi$ & $-\varrho$ & 0 \\ \hline \label{IIIf}
$0$ & $0$& 0 & $p$
 \end{tabular}\right).
\ee
All four Lorentz-invariant eigenvalues are zero if and only if $\varrho=p=0$. In this case the nonzero eigenvalues of the matrix $T^{\ha\hat{b}}$ are
\be
\tilde\lambda_{1,2}=\pm\sqrt{2}\varphi.
\ee

\section{Apparent horizon in the outgoing\\ Vaidya metric}           \label{aC}
 On a hypersurface $v=\mathrm{const}$ we introduce the surface coordinates $(\breve{r},\theta,\phi)$ where the bulk coordinate $r$ is expressed in terms of the coordinates $\breve{r}$ and $\theta$ as
 \be
 r=\breve{r}+z(\theta),
 \ee
 for some function $z$. Locating the apparent horizon $r_\sg(v,\theta)$ is then expressed as a problem of finding the function $z(\theta)$ such that $r_\sg=r_0(v)+z(\theta)$. While the function $z$ also depends on $v$,
  it does not affect the derivation below and this dependence is omitted.

Two spacelike vectors that are tangent to the surface $\breve{r}=\mathrm{const}$ are
\be
 b^\mu_\theta=z'(\theta)\delta^\mu_r+\delta^\mu_\theta, \qquad b^\mu_\psi=\delta^\mu_\psi.    \label{veb}
 \ee
 We obtain the outward- and inward-pointing future-directed null vectors $l^+\equiv l$
  and $l^-\equiv N$ by using the orthogonality condition $l^\pm_\mu b^\mu_{A}=0$.  Before the rescaling $l^v=1$ and the  normalization $N\cdot l=-1$  the two null vectors are given by
 \be
 l^\pm_\mu\propto(-1,\ell_\pm , - \ell_\pm z'(\theta),0). \label{C3}
 \ee
The two values of $\ell_\pm$ are obtained from the null condition $l^\pm\cdot l^\pm=0$,
 \begin{align}
 \ell_\pm &=\frac{1}{\Delta+ z'^2}\Big(r^2+a^2 \nonumber \\ &\pm\sqrt{2 a^2 r M \sin ^2 \theta+\rho^2 \left(a^2+r^2\right)-a^2 z'^2\sin^2\theta}\,\Big).
 \end{align}
 After setting $l^v=1$  the leading order components of the future-directed outward-pointing null vector orthogonal to the two-surface $r=r_0+z(\theta)$ are
 \begin{align}
l^v &=1, \\
 l^r&=\frac{(r_0^2-a^2)z'}{2r_0(r_0^2+a^2)},\\
 l^\theta  &=  -\frac{z'}{r_0^2+a^2}, \\
 l^\psi &=\frac{a}{r_0^2+a^2}\nonumber \\
&~+\frac{a(a^4-7a^2 r_0^2-10r_0^4-a^2(r_0^2-a^2)\cos2\theta)z}{4r_0(r_0^2+a^2)},
 \end{align}
  where we assume that $|z|\ll r_0$ and $|z'|\ll r_0$.

We now consider the change  in the two-dimensional  area  after one infinitesimal step $\delta\lambda$ of the evolution $x^\mu_\mathrm{in}\to  x^\mu_\mathrm{fn}$, where
\begin{align}
&x^\mu_\mathrm{in}=(v,r_0+z(\theta),\theta,0),    \\
& x^\mu_\mathrm{fn}=x^\mu_\mathrm{in}+l^\mu(x^\mu_\mathrm{in}) d\lambda,
\end{align} and $\lambda$ is the affine parameter.

 The  determinant of the two-dimensional metric $\sigma_{AB}$ is given by Eq.~\eqref{met2d}. To obtain the initial area the Kerr-Vaidya metric is evaluated   at $x_\mathrm{in}$ and the
vectors  $b_A$ are given by Eq.~\eqref{veb}.   To calculate the final area we evaluate the four-dimensional metric at the point $x_\mathrm{fn}$. In addition,  since the  points $x_\psi$ and  $x_\theta$ that
 are defined by Eq.~\eqref{2point}  evolve  with the vectors 
 \be 
 l^\mu(x_\psi)=l^\mu(x_\mathrm{in}),
 \ee
 and
 \be
 l^\mu(x_\theta)\approx l^\mu(x_\mathrm{in})+\pad_\theta l^\mu(x_\mathrm{in})\delta\theta, 
 \ee
respectively,
 % while the comoving coordinates $\theta$ and $\psi$ remain constant,
 the cross section tangents evolve as
 \be
 b^\mu_\psi \to  b^\mu_\psi, \qquad b^\mu_\theta \to  b^\mu_\theta+\pad_\theta l^\mu(x_\mathrm{in})d\lambda.
 \ee
The area differential $d\delta A\propto (d\sqrt\sigma/d\lambda)d\lambda$  is evaluated by subtracting $\sqrt{\sigma(x_\mathrm{in})}$ from the first-order
expansion in $d\lambda$ of $\sqrt{\sigma(x_\mathrm{fn})}$. The
desired Eq.~\eqref{odeZ} is obtained by setting $d\delta A=0$.

An alternative derivation is based on extending the vector field $l^\mu$ from the hypersurface  $v=\mathrm{const}$ to the bulk in such a way that the new field $\mathring{l}^\mu$
satisfies the geodesic equation $\mathring{l}^\mu_{;\nu}\mathring{l}^\nu=0$. In fact, this needs to be done only on the hypersurface itself, where it is realized by setting
\be
\mathring{l}^\mu\defeq l^\mu,    \ee
and thus $\mathring{l}^0=l^0=1$,
\be
 \qquad \mathring{l}^\mu_{;m}\defeq l^\mu_{\,;m},
 \ee
 for $m=1,2,3$, and setting the covariant derivative over $v$ as
 \be
 \mathring{l}^\mu_{\,;0}\defeq -l^\mu_{\,;m} l^m,
\ee
For the affinely parametrized geodesic congruence $\vartheta=\mathring{l}^\mu_{\,;\mu}$, and Eq.~\eqref{odeZ}  follows from
 \be
\vartheta=-l^0_{\,;m} l^m+l^m_{;m}=0.
 \ee
%%%%%%%%%%%%%%%%%%%%%%%%%%%%%%%%%%%%%%%%%%%%

\end{document}